\documentclass{aa}
\usepackage{graphicx}
\usepackage{txfonts}
\usepackage{natbib}
\usepackage{color}

\bibpunct{(}{)}{;}{a}{}{,}

\begin{document}

\title{Effects of a solid surface on jet formation around neutron stars}

\author{Matthias Stute\inst{1} \and Jos$\acute {\rm e}$ Gracia\inst{2} \and 
Max Camenzind\inst{1}}
\institute{Landessternwarte Heidelberg, K\"onigsstuhl, D-69117 Heidelberg, 
Germany \and  Section of Astrophysics, Astronomy \& Mechanics, Department of 
Physics, University of Athens, Panepistimiopolis, 157 84 Zografos, Athens, 
Greece}

\offprints{Matthias Stute, \email{M.Stute@lsw.uni-heidelberg.de}}

\date{Received 25 November 2004 / Accepted 03 February 2005}

\abstract{
We present two numerical simulations of an accretion flow
from a rotating torus onto a compact object with and without a 
solid surface -- representing a neutron star and a black hole -- 
and investigate its influence on the process of jet formation. We report the 
emergence of an additional ejection component, launched by thermal pressure 
inside a boundary layer (BL) around the neutron star and examine its 
structure. Finally, we suggest improvements for future models.
}

\maketitle

\section{Introduction}

Although jets are ubiquitous phenomena in many different astrophysical 
objects, their formation is relatively unclear. We find jets in  young stellar 
objects where they are driven by protostars, in symbiotic stars (white dwarfs),
X-ray binaries (neutron stars and stellar mass black holes) and active 
galactic nuclei (supermassive black holes). The mass loss rate of all jets is 
found to be connected to the mass accretion rate of the underlying disk found 
in most objects \citep[e.g.][]{Liv97}. Therefore the necessary components seem 
to be well known and common to all objects. 

In jet formation models presented so far, the magnetic field seems to play a 
key role. The first analytical work studying magneto-centrifugal acceleration 
along magnetic field lines threading an accretion disk was done by 
\citet{BlP82}. They have shown braking of matter in azimuthal direction 
inside the disk and its acceleration above the disk surface by the poloidal 
magnetic field components. Toroidal components of the magnetic field then 
collimate the flow. Numerous semi-analytic models extended the work of 
\citet{BlP82}, which either were restricted to self-similar solutions and 
their geometric limitations \citep[e.g.][]{PuN86,VlT98,VlT99,FeC04} or 
suggested non-self-similar solutions \citep[e.g.][]{Cam90,PeP92,BrC00}.

Another approach is to use time-dependant numerical MHD simulations to 
investigate the formation and collimation of jets. In most models, however, 
a polytropic equilibrium accretion disk was regarded as a boundary condition 
\citep[e.g.][]{KLB99,KLB04,ALK,GBW99}. The magnetic feedback on the disk 
structure is therefore not calculated self-consistently. Only in recent years 
were the first simulations presented including the accretion disk 
self-consistently in the calculations of jet formation 
\citep[e.g.][]{CaK02,CaK04,KMS04}. 

\citet{Pri89} proposed the idea of jet formation in the BL region and
also in his model strong magnetic fields were driving the outflows. 
\citet{Tor84} first assumed the liberation of energy in BL shocks 
to drive winds by thermal pressure. \citet{Liv99} pointed out that an 
additional source of energy beside the magnetic one is needed to power jets. 
\citet{ToG92} performed numerical simulations and found mass ejection only 
when they had {\em not} taken radiative cooling into account. However, 
their simulations were only one-dimensional for calculating the vertical 
structure of the BL. New examinations and modifications of this possibility of 
accelerating plasma close to the central object were done by \citet{SoR03} 
involving SPLASHs ({\em SPatiotemporal Localized Accretion 
SHocks}) in the BL. Locally heated bubbles expand, merge, and accelerate 
plasma to higher velocities than the local escape velocity. This scenario was 
introduced in analytic estimates. Now the numerical treatment needs to be 
improved using {\em multi-}dimensional simulations -- at first purely 
hydrodynamical ones, which are presented in this paper.

In Sect. \ref{sec_models}, we present our numerical simulations of an 
accretion flow onto a neutron star with a solid surface and onto 
an accreting black hole without one, with which we investigate its effects on 
the jet formation process (Sect. \ref{sec_eff}). Sections \ref{sec_acc} 
and \ref{sec_ejcomp} examine the structure of the accretion flow and of the 
additional ejection component, while a discussion follows in Sect. \ref{disc}.

\section{The numerical models} \label{sec_models}

In the following, we describe our computer code with equations, the model 
geometry, and the parameters. 

\subsection{The computer code}

With the code {\em NIRVANA} \citep{Zie98,Zie99} we solve the following set 
of differential equations of ideal non-relativistic magnetohydrodynamics 
\begin{eqnarray} \label{MHDequations}
\frac{\partial\,\rho}{\partial\,t} + \nabla\,(\rho\,{\bf v}) &=& 0 \nonumber \\
\frac{\partial\,(\rho\,{\bf v})}{\partial\,t} + \nabla\,
(\rho\,{\bf v}\otimes{\bf v}) &=& - \nabla\,p + \frac{1}{\mu}\,(\nabla\times
{\bf B})\times{\bf B} - \rho\,\nabla\Phi \nonumber \\
\frac{\partial\,e}{\partial\,t} + \nabla\,(e\,{\bf v}) &=& - p\,\nabla\,{\bf v}
\nonumber  \\
p &=& (\gamma - 1)\,e \nonumber \\
\frac{\partial\,{\bf B}}{\partial\,t} &=& \nabla\times({\bf v}\times{\bf B}) \\
\end{eqnarray}
with density $\rho$, velocity ${\bf v}$, internal energy $e$, pressure $p$, 
magnetic field ${\bf B}$, gravitational potential $\Phi$, magnetic 
permeability $\mu$, and adiabatic constant $\gamma$.

A set of common boundary conditions, including inflow, outflow (open), mirror 
and anti-mirror and rotational symmetry, all with their usual meanings, has 
already been defined in {\em NIRVANA}. The code and its boundary conditions 
were tested in many simulations \citep[][and references therein]{Zie98,Zie99}.

\subsection{Initial conditions}

Besides taking the more or less standard disk as initial condition, another 
approach is to begin with a rotating torus inside the computational domain. 
One advantage of this setup is that all material is already inside the domain 
initially, so no matter source has to be implemented on the boundaries. 
In this case, we could use the standard boundary conditions of {\em NIRVANA}.

Starting with the static hydrodynamics equations, from the momentum equation 
follows
\begin{equation} \label{statMHD}
\frac{1}{\rho}\,\nabla\,p = -\,\nabla\,\Phi + \frac{l^2}{r^3}\,\nabla\,r ,
\end{equation}
where $r$ is the cylindrical radius. With a polytropic equation of state and 
the identity
\begin{equation}
\nabla\,(\frac{p}{\rho}) = \frac{\gamma - 1}{\gamma}\,\frac{1}{\rho}\,
\nabla\,p 
\end{equation}
Eq. (\ref{statMHD}) can be integrated to
\begin{equation}
\frac{\gamma}{\gamma - 1}\,\frac{p}{\rho} + \Phi + \int_{r}^{\infty}\,
dr'\,\frac{l (r')^2}{r'^3} = W_{0} .
\end{equation}
Under the assumption of constant $l$, the integral can be solved to
\begin{equation}
\frac{\gamma}{\gamma - 1}\,\frac{p}{\rho} + \Phi + \frac{1}{2}\,\frac{l^2}{r^2}
 = W_{0} 
\end{equation}
with which the density is then
\begin{equation}
\rho = \left( \frac{1}{\kappa}\,\frac{\gamma - 1}{\gamma}\,\left[ W_{0} - 
\Phi - \frac{1}{2}\,\frac{l^2}{r^2} \right] \right)^{1/(\gamma - 1)} .
\end{equation}
The angular momentum of the torus is then dependent on its radial position 
$R_0$ as
\begin{equation} \label{angmom}
l ( R_0 ) = \sqrt{G\,M\,R_0}\,\frac{R_0}{R_0 - 2\,R_{\rm g}} ,
\end{equation}
if a pseudo-Newtonian gravitational potential \citep{PaW80}
\begin{equation}
\Phi = - \frac{G\,M}{R - 2\,R_{\rm g}} 
\end{equation}
is chosen. $R$ is the spherical radius, and all distances are now given in 
units of $R_{\rm g}$. The corresponding time scale $t_0$ is then the inverse 
of the Keplerian period $t_0 \approx 10^{-4}$ s in our simulations. The 
density maximum of the torus was positioned to 8 $R_{\rm g}$. 

Inside the torus, the velocity components are then 
\begin{equation}
v_{\phi} = \frac{l ( R_0 )}{r} \qquad v_{r} = v_{\theta} = 0 .
\end{equation}
Outside the torus, Keplerian rotation is assumed 
\begin{equation}
v_{\phi} = \Omega_{\rm K}\,r \qquad v_{r} = v_{\theta} = 0 
\end{equation}
with 
\begin{equation}
\Omega_{\rm K} = \sqrt{G\,M\,R}\,\frac{R}{R - 2\,R_{\rm g}}\,\frac{1}{r^2} .
\end{equation}

To initialise the magnetic field and to assure that its divergence vanishes, 
we calculate the magnetic field components from the vector potential $\vec A$ 
defined as
\begin{equation}
\vec B = \nabla\,\times\,\vec A.
\end{equation}
The only considered component of $\vec A$ is set identical to the internal 
energy 
\begin{equation}
A_{\phi} = e .
\end{equation}
The initial magnetic field lines are then along isocontours of energy and 
density, as we used a polytropic equation of state during initialisation. This 
results in the following magnetic field components
\begin{equation}
B_{R} = \frac{1}{R\,\sin{\theta}}\,\frac{\partial}{\partial\,\theta} \, 
( \sin{\theta}\,e ) \quad B_{\theta} = - \frac{1}{R}\,
\frac{\partial}{\partial\,R}\, ( R\,e ) \quad B_{\phi} = 0 
\end{equation}
in spherical coordinates, or
\begin{equation}
B_{r} = - \frac{\partial}{\partial\,z} e \qquad B_{z} = \frac{1}{r}\,
\frac{\partial}{\partial\,r}\, ( r\,e ) \qquad B_{\phi} = 0 
\end{equation}
in cylindrical coordinates. Afterwards, these components were 
scaled to achieve an assumed plasma $\beta = p_{\rm gas} / p_{\rm mag} = 
10^{3}$. Note that no global external magnetic field was implemented.

We performed two simulations with different boundary conditions at the inner 
radial coordinate, one with open boundaries (Run A) -- describing a black hole 
-- and one with anti-mirror conditions to model the solid surface of the
central object (Run B) -- a neutron star. The other boundary conditions are 
open ones at the outer radial coordinate, rotation symmetry and 
anti-mirror symmetry at the inner and outer poloidal ($\theta$-) coordinate, 
respectively, to simulate the rotational axis and the equatorial plane, and 
periodic conditions at both azimuthal ($\phi$-) boundaries. The aim of 
these simulations was to investigate the influence of the solid surface. A 
third simulation (Run C) was performed similar to Run A, but in cylindrical 
coordinates to investigate the influence of different coordinate systems. Here 
the cylindrical radius starts at $2\, R_{\rm g}$; i.e. a cylinder along the 
axis is cut out and the inner radial boundary is chosen to be open.

\section{The effect of a solid surface on the jet formation process} 
\label{sec_eff}

\subsection{The accretion and ejection components}

\begin{figure}
  \resizebox{\hsize}{!}{\includegraphics{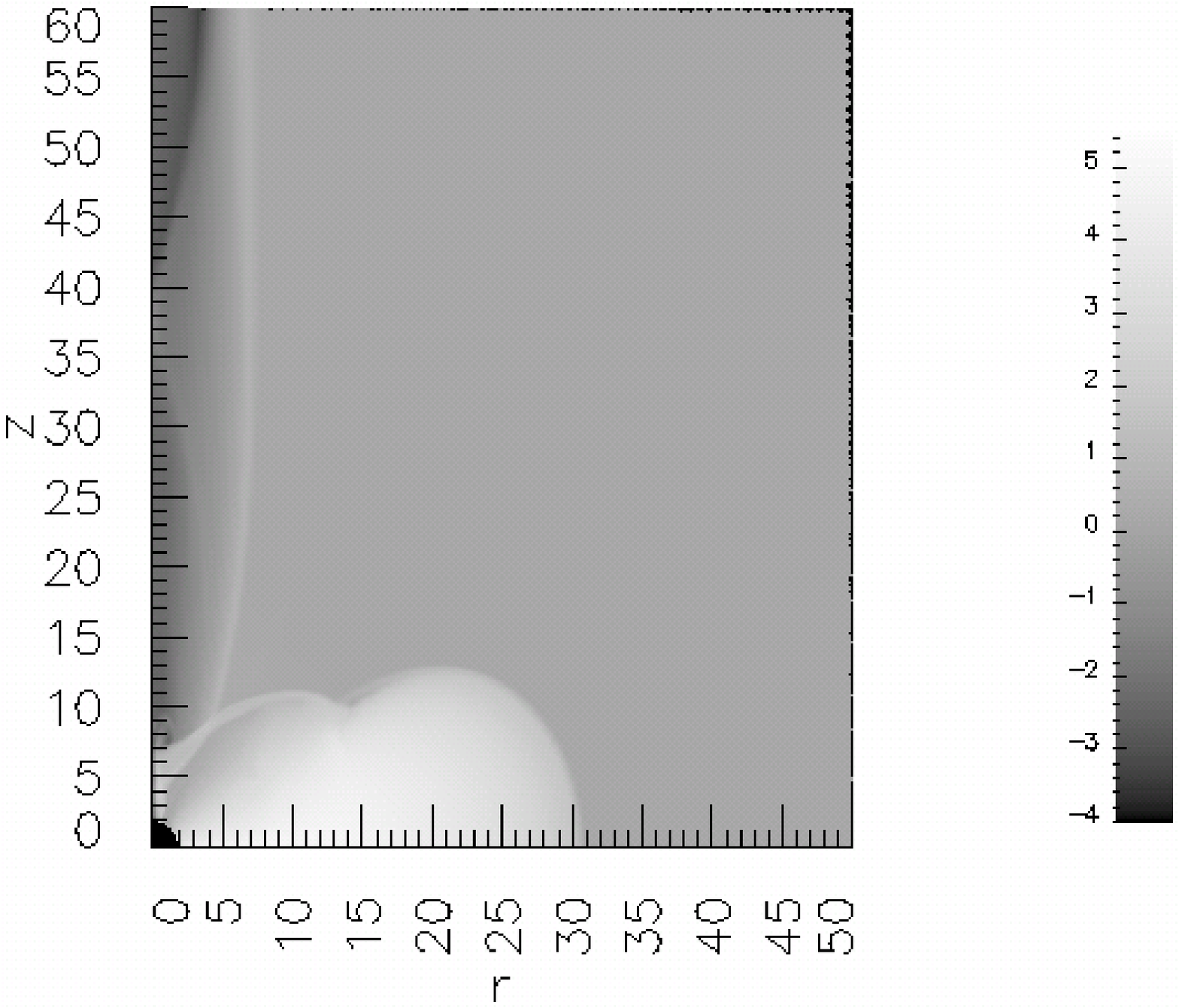}}
  
  \resizebox{\hsize}{!}{\includegraphics{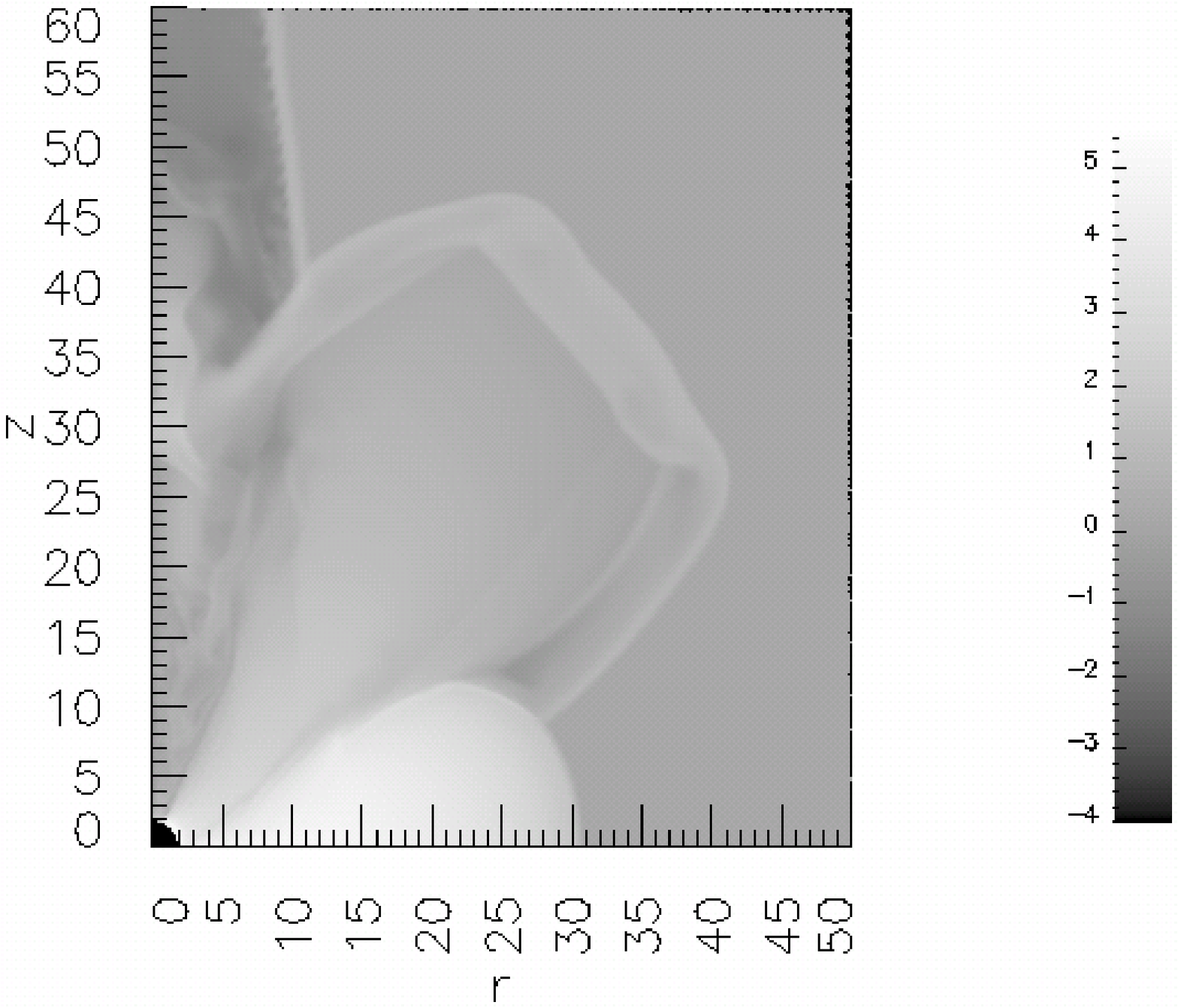}}
  
  \resizebox{\hsize}{!}{\includegraphics{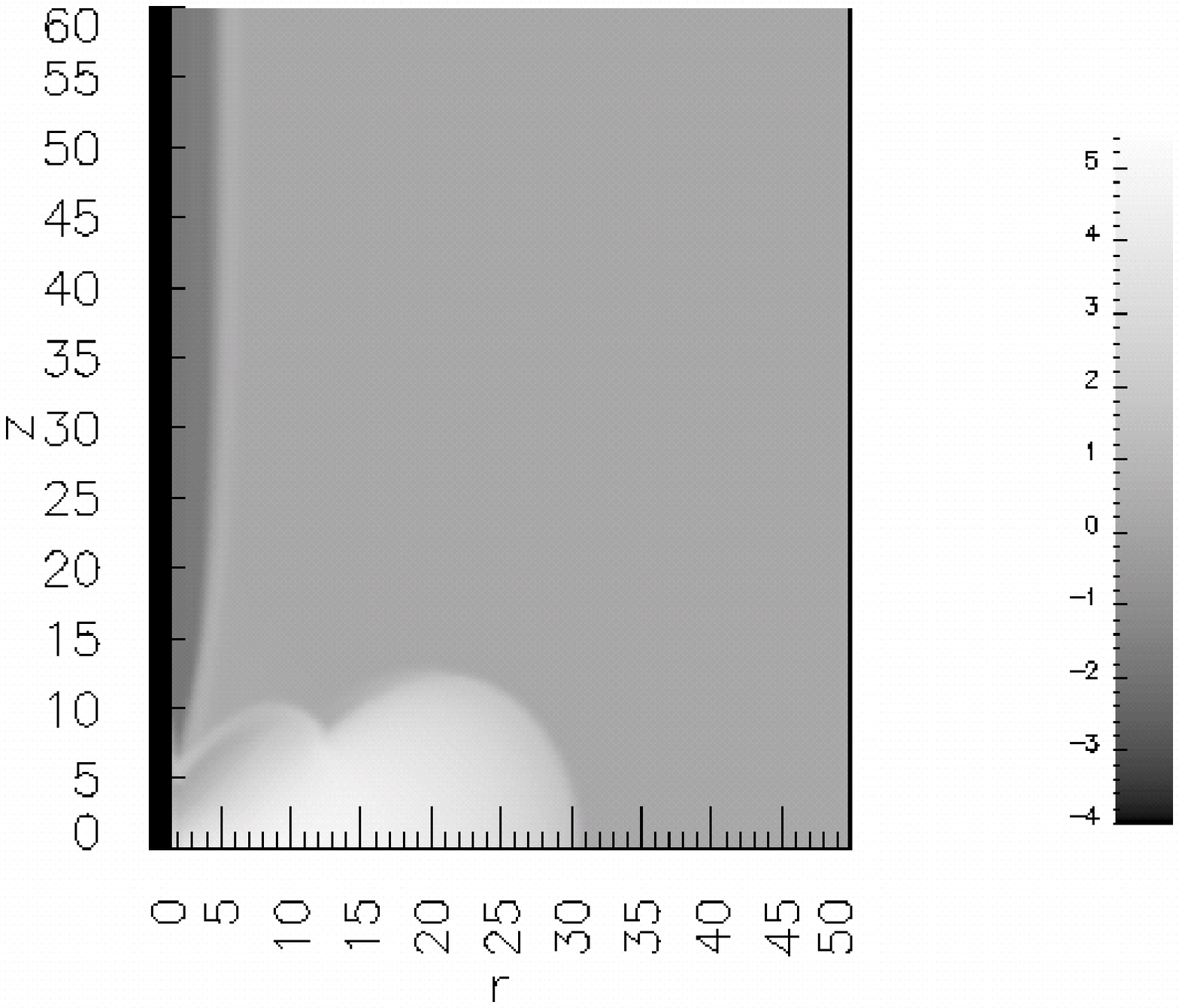}}
  \caption{Plots of the logarithm of density of Run A 
    (top), Run B (middle) and Run C (bottom) after 45 $t_0$; the
    coordinates are in units of $R_{\rm g}$}
  \label{runABC}
\end{figure}

In Fig. \ref{runABC}, the logarithm of density is plotted for the three runs 
after 45 $t_0$. One can clearly see that activity is triggered much faster in 
Run B with the solid surface boundary.

In Run A accretion sets in immediately because of the magneto-rotational 
instability \citep{BaH} inside the torus. After almost one revolution of the 
torus, it comes in contact with the central object, which leads to a 
re-distribution of matter in the now established disk. The inner part of the 
disk puffs up creating an expanding bubble. Along the rotation axis, a funnel 
-- i.e. a region evacuated by centrifugal forces to densities that are three 
orders of magnitude below the surrounding -- is created and its cross-section 
grows with time. 

In Run B the accretion process also starts immediately and the funnel is 
created. Additionally, a boundary layer with a radial extent of one fifth of 
the inner radius forms on the surface of the central object. Due to its 
high pressure, a flare occurs. This flare bubble, now created by the boundary 
layer and not by the accretion disk, expands along the surface of the torus 
with high velocities and high density. The density contrast of the 
bubble is $\eta = 10^{2}$, i.e. the bubble is overdense. Expansion of this 
bubble is powered by a continuous high pressure flow from the boundary layer. 
The expansion direction of the bubble is along a latitude of around 
40-50$^{\circ}$. If an external magnetic field were present, perhaps the 
flow would bend towards the axis.

Only in Run C, a high velocity component emanates out of the spherical 
expansion inside the swept-out funnel along the axis after 64.5 $t_0$. This 
funnel jet has a head velocity between $c/3$ and $c$ and a density contrast of 
$\eta = 10^{-2}$. As this is only seen in Run C and not in Run A, the open 
boundary might create this feature. 

\begin{figure}
  \resizebox{\hsize}{!}{\includegraphics{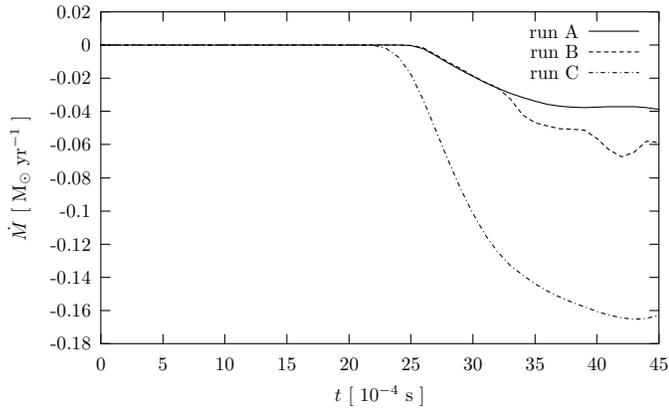}}
  \caption{Time evolution of accretion rate 
    $\dot M = -4\,\pi\,\rho\,v_{r}\,H$ in the 
    equatorial plane at $r = 4\,R_{\rm g}$. In Run C the 
    accretion rate is always higher than in the other simulations, 
    perhaps due to the boundary conditions. After 21 $t_0$, the flare coming 
    from the boundary layer destabilizes the accretion flow and increases its 
    rate in Run B.}
  \label{mdot_time}
\end{figure}

In Fig. \ref{mdot_time}, the time evolution of the accretion rate in the 
equatorial plane at $r = 4\,R_{\rm g}$ is plotted. It is 
calculated as $\dot M = -4\,\pi\,\rho\,v_{r}\,H$ with height $H$ of the 
disk. The qualitative behavior seems to be equal in all runs, but in Run 
C the accretion rate is always higher. It is possible that the open boundary 
near the axis causes a global radial inflow which is superimposed on the 
accretion rate visible in Runs A and B. The accretion rate in the Runs A and B
seems to be equal until about 21 $t_0$; i.e. no causal connection is 
established between the boundary layer and the accretion flow itself until 
then. After that point, the flare coming from the boundary layer destabilizes 
the accretion flow and increases its rate.

\subsection{Jet emission efficiency}

\begin{figure}
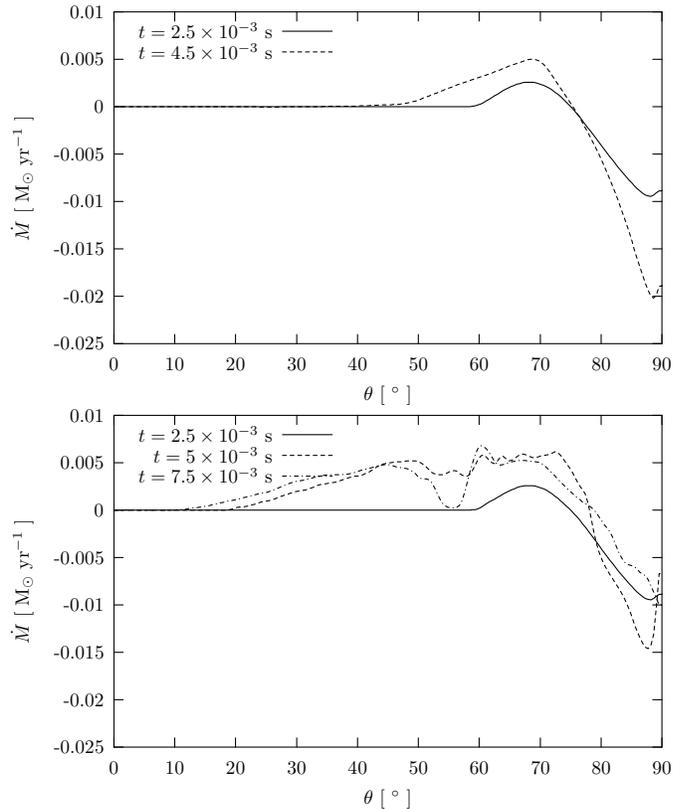

  \resizebox{\hsize}{!}{\includegraphics{mdot_theta_runA.eps}}

  \resizebox{\hsize}{!}{\includegraphics{mdot_theta_runB.eps}}
  \caption{Poloidal slice at $r = 10\,R_{\rm g}$ of the mass accretion/outflow 
    rate $\dot M = -4\,\pi\,\rho\,v_{r}\,H$ of Runs A (top) 
    and B (bottom) at different times. The emergence of an 
    additional outflow component is visible at about 40$^{\circ}$.}
  \label{mdot_theta}
\end{figure}

In Fig. \ref{mdot_theta}, a poloidal slice at $r = 10\,R_{\rm g}$ of the 
mass accretion/outflow rate -- again calculated as 
$\dot M = -4\,\pi\,\rho\,v_{r}\,H$ -- is plotted at different times for Run A 
and for Run B. One can clearly see the different accretion and ejection 
components in this simulation. 

\begin{figure}
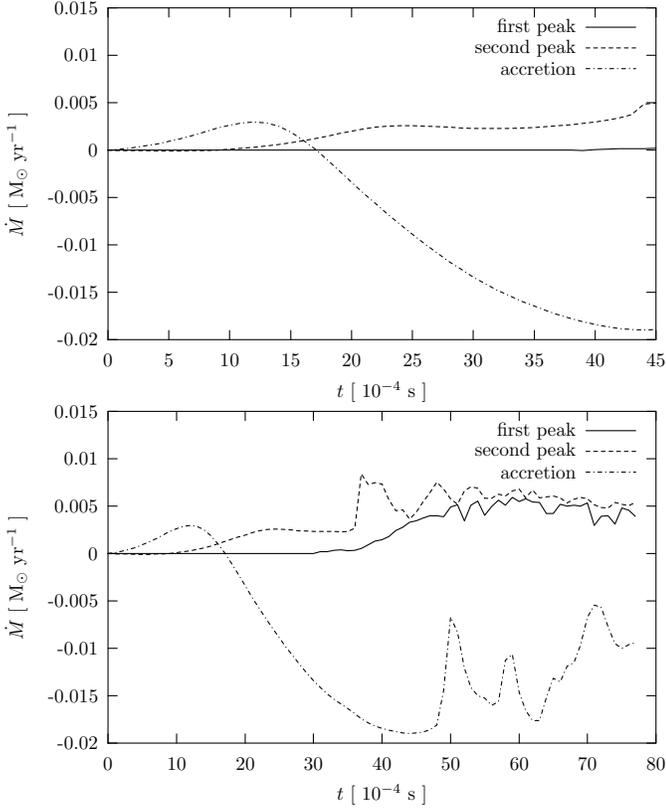

  \resizebox{\hsize}{!}{\includegraphics{mdot_peak_runA.eps}}

  \resizebox{\hsize}{!}{\includegraphics{mdot_peak_runB.eps}}
  \caption{Time evolution of the ejection peaks and the accretion disk in Run 
    A (top) and Run B (bottom). While the accretion/outflow rates 
    seem to reach an asymptotic value in Run A, they are highly variable in 
    Run B due to the flary conditions inside the BL.}
  \label{mdot_peak}
\end{figure}

The first peak at about 40$^{\circ}$ represents the flares created by the 
boundary layer in Run B and has no corresponding feature in Run A. The second 
peak at about 70$^{\circ}$ is identical to an outflow along the surface 
of the torus, which is common in all runs. At larger values of $\theta$, the 
accretion flow can be seen. At this distance of $10\,R_{\rm g}$ the 
accretion rate is higher in Run A than in Run B, while at closer distance the 
behavior is reversed (Fig. \ref{mdot_time}). The accretion rate in all peaks 
is highly time dependent in Run B, while it seems to reach an 
asymptotic value in Run A (Fig. \ref{mdot_peak}), which results from the flary 
conditions inside the BL. 

\begin{figure}
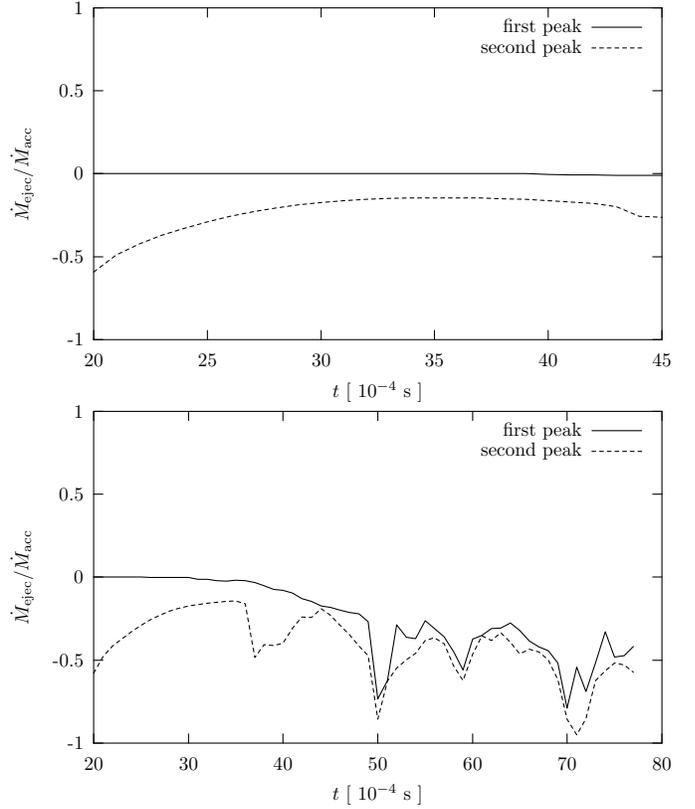

  \resizebox{\hsize}{!}{\includegraphics{efficiency_runA.eps}}
  \resizebox{\hsize}{!}{\includegraphics{efficiency_runB.eps}}
  \caption{Efficiency of the ejection mechanism, calculated as 
      the ratio of the outflow rate in the ejection peak to the accretion rate 
      in the equatorial plane plotted for Runs A (top) and B (bottom)}
  \label{ej_ac_eff}
\end{figure}

Using the mass accretion rate in the equatorial plane and the mass ejection 
rate of the two peaks, one can calculate an ejection efficiency of the system 
as the ratio of both rates. This efficiency is plotted in Fig. \ref{ej_ac_eff}.
After an initial phase of global ejection in the equatorial plane until about 
14 $t_0$, the mass fraction outflowing along the torus surface (second peak) 
compared to that being accreted is almost constant between 25--30 \% in Run A, 
while oscillating around a mean of about 50 \% in Run B. In Run A, the 
efficiency of ejection in the first peak is only one percent and the ejection 
peak is not really present. In Run B, however, the ejection efficiency of the 
first peak is comparable to that of the second peak, i.e. also in the range 
between 40-50 \%. Therefore almost the whole accreted matter is 
ejected out of the central region.

\section{Structure of the accretion flow} \label{sec_acc}

\begin{figure}
  \resizebox{\hsize}{!}{\includegraphics{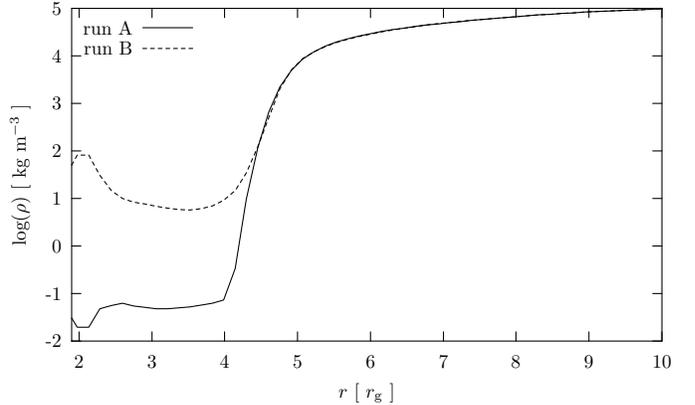}}
  \caption{Structure of the accretion flow after 14 $t_0$: 
    density. The density increases in the BL by more than an 
    order of magnitude with respect to the accretion flow, in Run A, the open 
    boundary representing the inner sink arranges for a density decrease by a 
    factor of 4-5.}
  \label{disk_struct_den}
\end{figure}

\begin{figure}
  \resizebox{\hsize}{!}{\includegraphics{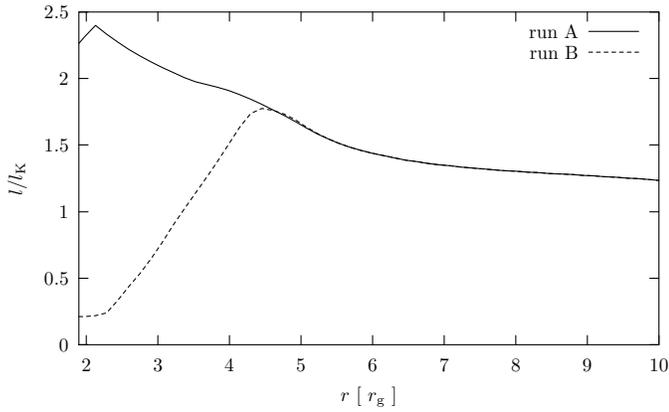}}
  \caption{Structure of the accretion flow after 14 $t_0$: 
    ratio of angular momentum to Keplerian angular momentum. In Run B the 
    accretion flow is always rotating super-Keplerian and the angular 
    momentum of the flow in Run A drops below the Keplerian angular momentum 
    for distances smaller than about 3 $R_{\rm g}$.}
  \label{disk_struct_l}
\end{figure}

\begin{figure}
  \resizebox{\hsize}{!}{\includegraphics{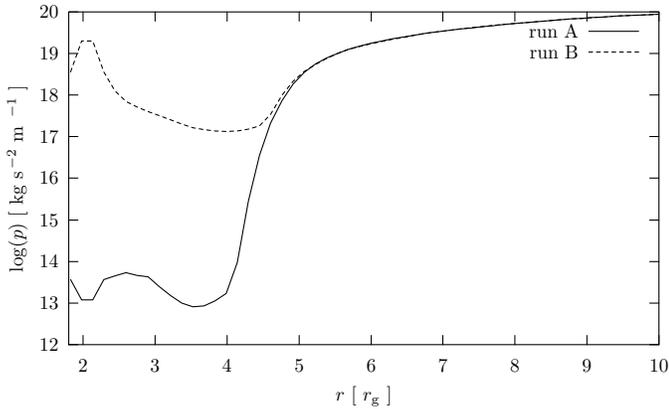}}
  \caption{Structure of the accretion flow after 14 $t_0$: 
    pressure. The thermal pressure is enhanced in Run B inside 
    the BL by six orders of magnitude with respect to Run A.}
  \label{disk_struct_p}
\end{figure}

\begin{figure}
  \resizebox{\hsize}{!}{\includegraphics{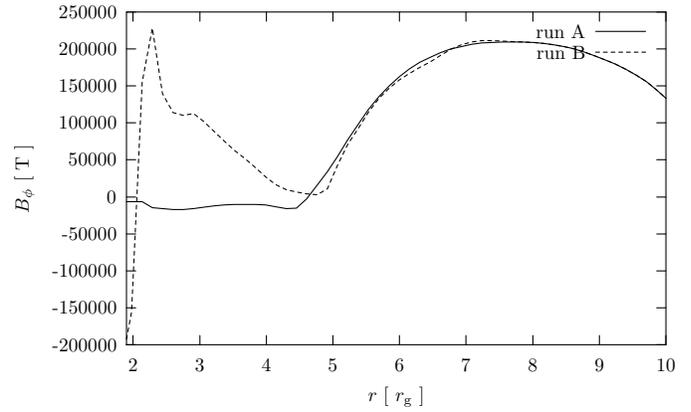}}
  \caption{Structure of the accretion flow after 14 $t_0$: 
    azimuthal magnetic field component. The emergence of the 
    boundary layer also creates a peak in the magnetic field at its surface, 
    which is perhaps caused by compression in strong shocks inside it.}
  \label{disk_struct_bz}
\end{figure}

The next step is to investigate the structure of the accretion flow and to 
raise the question of influences of the emergence of the boundary layer. In 
Figs. \ref{disk_struct_den}-\ref{disk_struct_bz}, the main 
magnetohydrodynamical quantities of the flow are plotted.

\begin{figure}
  \resizebox{\hsize}{!}{\includegraphics{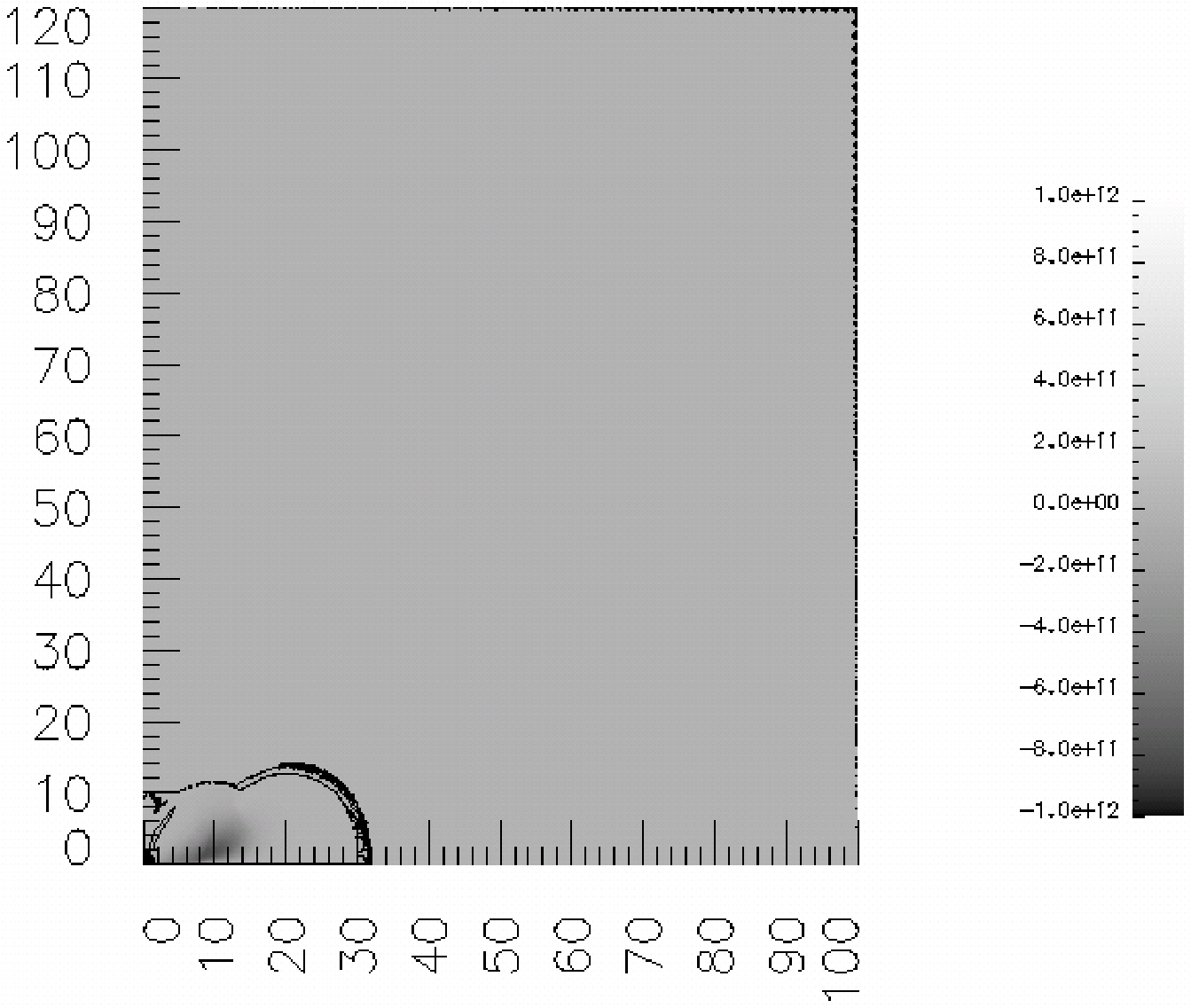}}
  
  \resizebox{\hsize}{!}{\includegraphics{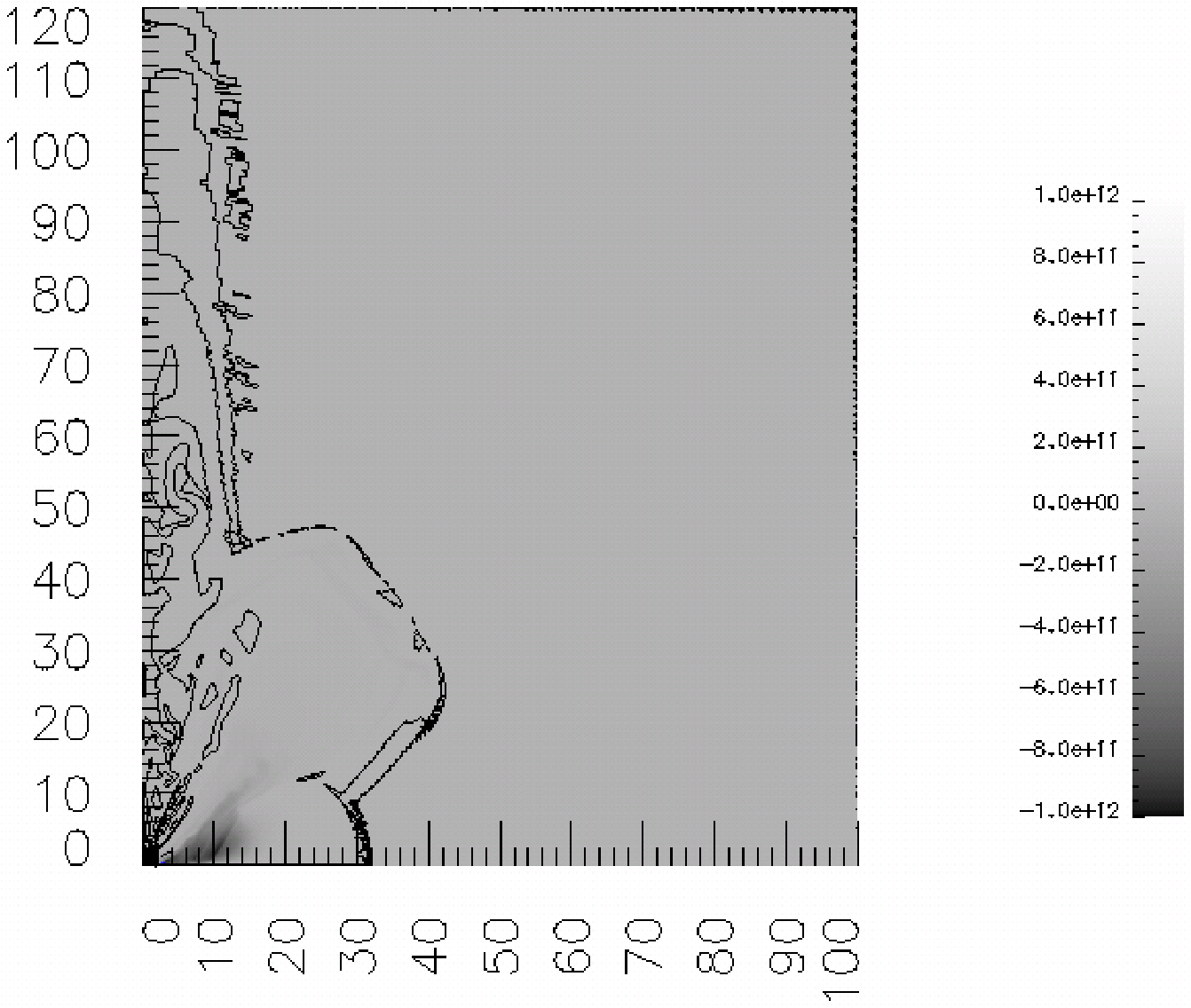}}
  
  \resizebox{\hsize}{!}{\includegraphics{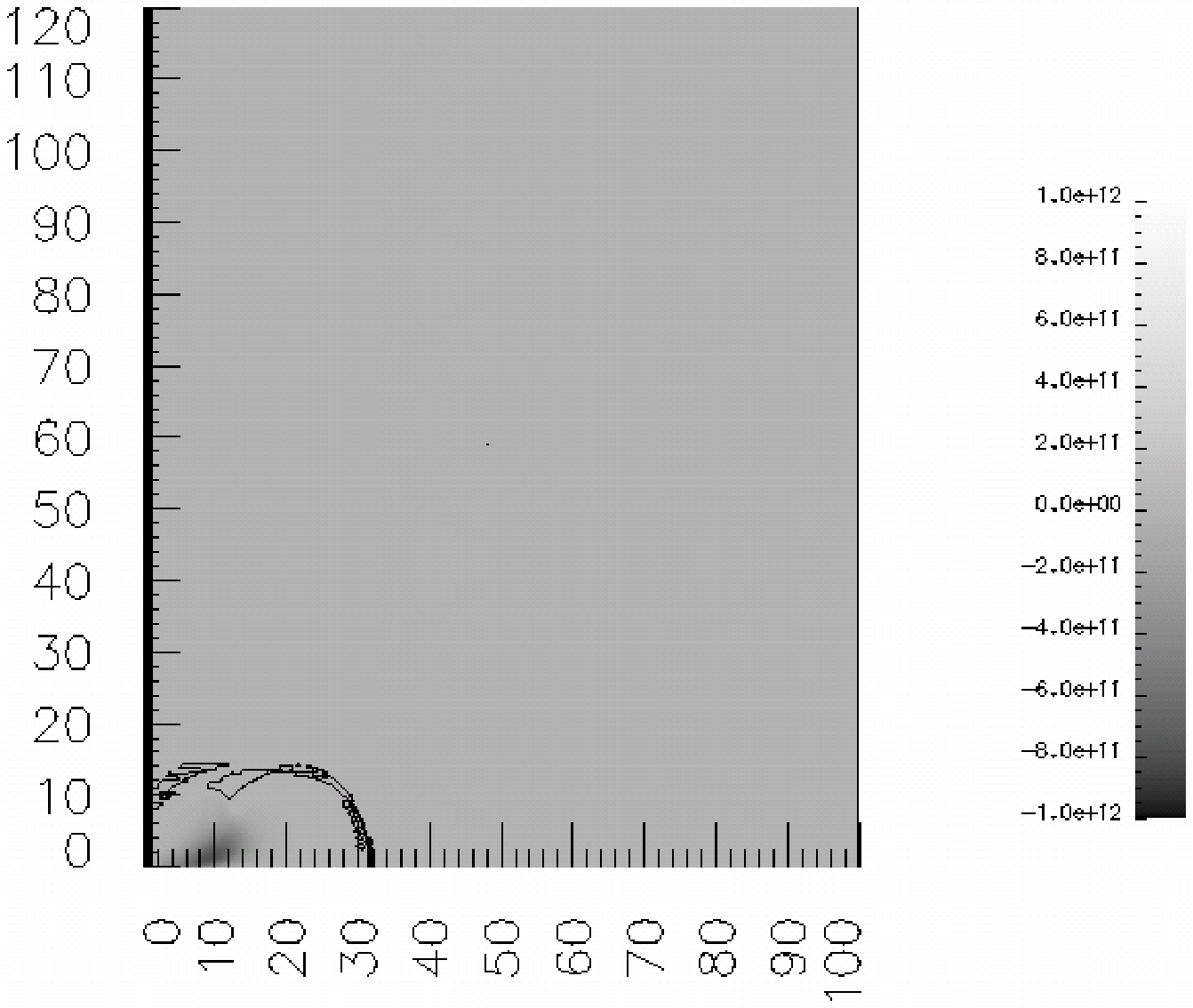}}
  \caption{Plots of the total poloidal current $r\,B_{\phi}$ after 31 $t_0$ 
    for Run A (top), Run B (middle) and Run C (bottom) with logarithmic 
    contour lines from $I = 0$ to $I = 10^{12}$ A}
  \label{current}
\end{figure}

In Fig. \ref{disk_struct_den},which shows the density distribution 
in the equatorial plane, the boundary layer between 2 and 2.4 $R_{\rm g}$ 
(1-1.2 radii of the central object) is, along with the torus, the most 
prominent feature in Run B. The density increases by more than an order of 
magnitude with respect to the accretion flow. In Run A, the open boundary 
representing the inner sink creates a density decrease by a factor of 4-5,
due to draining into the black hole. In contrast to Run B, in which the 
accretion flow is always rotating super-Keplerian, the angular momentum of the 
flow in Run A drops below the Keplerian angular momentum for distances smaller 
than about 3 $R_{\rm g}$ (Fig. \ref{disk_struct_l}). This is also an effect 
of the drag created by the black hole. The thermal pressure is enhanced in Run 
B inside the BL by six orders of magnitude compared to Run A (Fig. 
\ref{disk_struct_p}). The emergence of the boundary layer also creates a peak 
in the radial and azimuthal components of the magnetic field (Fig. 
\ref{disk_struct_bz}) at its surface, which is perhaps caused by compression 
in strong shocks inside it. This highly magnetised region extends to a 
distance of 60 $R_{\rm g}$ from the equatorial plane, leading to a positive 
total poloidal current instead of negative values near and inside the torus 
(Fig. \ref{current}), which then can also drive jets magnetically.

\section{Structure of the BL ejection component} \label{sec_ejcomp}

\begin{figure}
  \resizebox{\hsize}{!}{\includegraphics{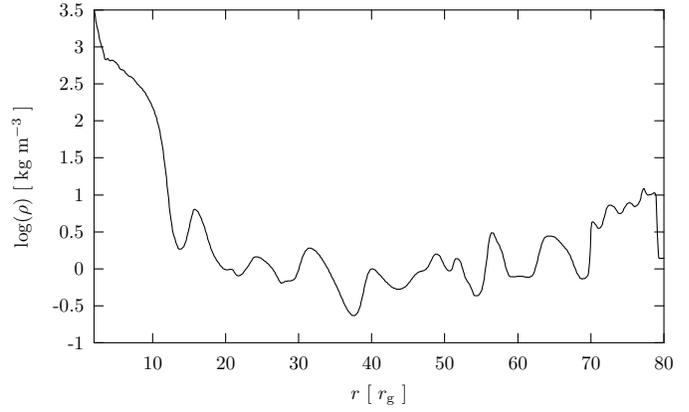}}
  \caption{Structure of the ejection component at 
    $\theta = 45^{\circ}$ after 54 $t_0$: density. Knots with 
    enhanced density are present.}
  \label{ejec_struct_den}
\end{figure}

\begin{figure}
  \resizebox{\hsize}{!}{\includegraphics{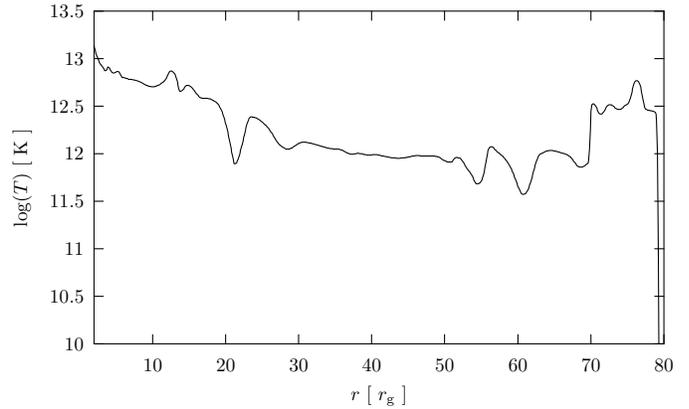}}
  \caption{Structure of the ejection component at 
    $\theta = 45^{\circ}$ after 54 $t_0$: temperature}
  \label{ejec_struct_T}
\end{figure}

In Figs. \ref{ejec_struct_den}-\ref{ejec_struct_T} the density and 
temperature along a slice through the ejection component at 
$\theta = 45^{\circ}$ are plotted.
The variability of the accretion and ejection rates leaves its mark in the 
form of a rich substructure in all variables, especially in density. 
One can distinguish knots with enhanced density, magnetic field components, 
and temperature. In the velocity components, the knots are  
modulated on a global deceleration of the flow. The jet is still highly 
transient and simulations with an extended computational domain have to show 
whether a steady outflow on larger scales can be established.

\section{Discussion} \label{disc}

We set up numerical simulations of a compact object with and without a
solid surface accreting matter from a rotating torus. They show an 
additional ejection component that could be collimated into a jet by a global 
magnetic field which was, however, omitted in our simulations. 
Another possibility for achieving collimation is an external pressure 
gradient, which was also not present in our simuations.

Our results seem to support the {\em SPLASH} scenario of \citet{SoR03}. We 
reported the emergence of a ejection component which is highly variable in 
agreement with their scenario. In our simulations a two-dimensional treatment 
was used, the full description in three dimensions could reveal new effects or 
details which we have to look for in new simulations. 

Another point is the amount of physics in our models. We used the equations of 
{\em ideal} MHD neglecting any cooling effects. Depending on the accretion 
rate, however, the boundary layer material can become optically thick, which 
has to be taken into account. The additional equations in the flux-limited 
diffusion ansatz can no longer be solved by {\em NIRVANA}, so that a new tool 
has to be used e.g. {\em FLASH}, and a new branch of simulations will be 
necessary. As cooling effects reduce the internal energy of the material and 
thereby also reduce its thermal pressure, it has to be shown, whether this 
new jet formation scenario is still reliable in two- or three-dimensional 
models. 

We have presented simulations with a set of model parameters appropriate for 
an accreting neutron star and an accreting black hole, respectively. The 
accretion rate in our simulations is, however, far too large for XRBs, but
would only be suitable for a gamma-ray burst. These are created by an 
overestimated density inside the rotating torus. Further simulations need to 
show whether this scenario still works at lower rates and will have to fix the 
value of a critical accretion rate, as stated in the analytic model by 
\citet{SoL}.

The equations of ideal MHD can theoretically be written in a non-dimensional 
form, if one uses a Newtonian gravitational potential instead of the 
pseudo-Newtonian one. Neglecting the latter, we could normalize all quantities 
to naturally arising combinations which depend on parameters of the central
object and carry over our results to other jet sources. However, additional 
simulations representing other classes of jet emitting objects will follow.

\begin{acknowledgements}
Parts of this work were supported by the Deutsche Forschungsgemeinschaft (DFG) 
and by the European Community's Research Training Networt RTN ENIGMA under 
contract HPRN-CT-2002-00231. We acknowledge the useful comments and 
suggestions by the anonymous referee.
\end{acknowledgements}

\end{document}